# Realization of High Frequency Bidirectional Transceiver (Bitx) Radio

Muhammad Firdaus[1], Sutrisno[1], and Tata Supriyadi[1]
[1]Department of Electrical Engineering, Politeknik Negeri Bandung, Indonesia

muhammadfirdaus.mufi@gmail.com, {sutrisno, tatasupriyadi}@polban.ac.id

*Abstract*—**In this paper, we realize the transmitter of high frequency (HF) bidirectional transceiver radio. A bidirectional transceiver (Bitx) consists of a transmitter and receiver, in which some common circuits are used together by the transmitter and receiver. In addition, Bitx radio uses a single conversion superheterodyne transmitter, which only shifts frequency once in a signal generation process. The measurement result shows that it works at the frequency of 6.8 MHz with 10 MHz BFO, 3,2 MHz VFO, and 50,3 dBm transmit power. The transmitting frequency is obtained from the difference between BFO and VFO oscillator. Bitx radio has a simple circuit and is designed with low-cost components, and it could be an alternative solution for communication in remote areas.**

## I. INTRODUCTION

A transceiver radio communication is a radio device that is responsible for being a sender and receiver used for communication purposes. In the early generations, the transmitter and receiver parts were designed separately, independently, and worked with their own functions. However, with the current development of radio communication, the two parts are integrated alternately in the same series [12]. However, the cost of factory-made radio transceivers with good quality and a wide range is still quite high, making it difficult for most people to buy. In addition to being quite expensive, radio transceivers have a complex circuit with separate transmitter and receiver functions even though they are in the same circuit. Hence, it is an obstacle for people trying to make radio transceivers at home. Therefore, we need an inexpensive radio transceiver with a simple circuit that can easily be made.

To overcome the above challenges, we aim to realize a high frequency (HF) bidirectional transceiver (Bitx) as a solution for amateur radio enthusiasts to construct inexpensive radio transceivers with better quality. Moreover, Bitx can be an alternative communication solution for remote areas lacking access to information or utilized during a disaster. Bitx20 from Mr. Ashar Farhan inspires the realization of our HF Bitx. Some of the parameters that must be met include to described as follow:
1. Produce the 7 MHz working frequency according to the HF frequency band
2. The output frequency of the BFO (Beat Frequency Oscillator) is expected to be 10 MHz because it uses a 10 MHz crystal.
3. The output frequency of VFO (Variable Frequency Oscillator) is expected to be 3 MHz to obtain a working frequency of 7 MHz
4. The oscillator, mixer, and filter circuit are designed bidirectional because they have the same function and can be used together

The rest of this paper is arranged as follows: we briefly explain the background knowledge and concept of bidirectional transceiver radio in Section 2. Section 3 explains the circuit and block diagram of the bidirectional transceiver radio. Then, we examine the results and considerations in Section 4. Finally, Section 5 concludes this paper.

## II. BACKGROUND

Bidirectional transceiver (Bitx) radio is a work created by Mr. Ashhar Farhan, an amateur radio enthusiast from India. The main goal of making Bitx is to design an inexpensive HF radio transceiver as a solution for Indian amateur radio enthusiasts who are, on average, below the marginal line and cannot afford to buy a factory-made radio that is quite expensive. Finally, in 2004, Farhan realized a Bidirectional radio SSB QPR Transceiver with 6 watts of power at a frequency of 14 MHz, later known as Bitx20. For more details on how Farhan made Bitx20, see [1]. A few years later, Bitx 20 has become an interesting thing for amateur radio fans worldwide, many of whom have succeeded in realizing and developing it. So do not be surprised if Bitx is said to be Open Radio because of its popularity. Interest in Bitx also spread to Indonesia; in 2008, an Indonesian radio amateur, Cholis Safrudin, succeeded in realizing Bitx in the 80-meter frequency band. Hence, it then prompted other amateur radio in Indonesia to try to design this cheap radio transceiver. For more details regarding the Bitx 80 version of Cholis Safrudin, you can read [5].

*A. Transceiver radio*

A radio communication transceiver is a radio device that also doubles as a radio receiver used for communication purposes. It consists of a sender (transmitter) and receiver (receiver), designed in an integrated manner. In the early generations, the transmitter and receiver parts were designed separately, were independent, and worked with their own functions. However, with the current development of radio communication, the two parts are integrated and function alternately in the same series [12].

The simple transmitter consists of an oscillator generating radio vibrations, and these vibrations, after being carried (modulated), are then converted by an antenna into radio waves and transmitted. It is known that sound waves have a low frequency (300 Hz – 3.4 kHz), so they cannot reach long distances even though the power is large enough, while radio waves with relatively small power can transmit information reaching thousands of kilometers [6]. For sound (information) to reach long distances, the sound is superimposed on radio waves resulting from radio vibration generators, which are called carrier waves. Then the carrier wave (carrier) will deliver the sound (information) to a distant destination. In that distant place, the radiated radio waves are received by the receiving radio antenna [9]. On the antenna side, radio waves in the form of electromagnetic waves are converted into electrical vibrations and enter the receiver. The carrier vibration is discarded in the radio receiver, and the sound vibrations are accommodated and raised through the speakers. Therefore, an audio vibration can reach a long distance with this modulation technique [7].

Sound vibrations enter the transmitter through the microphone, and the microphone output often needs to be amplified first with an audio amplifier called a microphone pre-amplifier so it can be superimposed on the carrier by the modulator. In order to increase the transmit power of a transmitter, the vibration of the oscillator [10] is amplified first with a radio frequency amplifier before being transmitted. Reinforcement can be done once and can also be done more than once. Transmitters not amplified are called single-stage transmitters, and those amplified once are called two-stage and so on [11].

*B. Transmission mode*

The transmission mode in the communication system is divided into three categories: simplex, half-duplex, and full-duplex transmission modes. Simplex is communication that only takes place in one direction, from the sending station to one or many receiving stations. Examples of this mode of communication are commercial radio broadcast communication systems and television broadcast systems. Half-duplex is communication that takes place in both directions alternately. At a time when station A sends information to station B, station B can only receive information from station A and cannot send information to station A. Likewise, when station B sends information to station A, station A can only receive information from station B and cannot send information to station B. These phenomena happens in SSB communications such as CB communications and Handy Talky [2]. Meanwhile, full-duplex is communication that can take place simultaneously in both directions, for example, wired telephone communication.

*C. Band frequency allocation*

Frequency is a limited natural resource, so the world organization ITU (International Telecommunication Union) makes the necessary arrangements to utilize these frequency resources together. These settings will mainly provide a situation where there will be no interference between equipment systems operated in this world because each system works at the frequency according to its designation [3]. The regulation related to the frequency band is stated in the RR (Radio Regulation) book issued by the IFRB (International Frequency Regulation Board), one of the administrative bodies under ITU [4].

The frequency allocation of radio waves is divided into several bands, as shown in Table 1. Table 1 describes the frequency range with the lower and upper limits of the frequency of the related band, containing the names of the bands and the use of frequency bands in the telecommunications system [8].

### III. REALIZATION OF HF BIDIRECTIONAL TRANSCEIVER RADIO

In this section, we explain the circuit and block diagram of the bidirectional transceiver radio. Radio bidirectional transceiver (Bitx) is generally divided into two main lines: the transmitter and receiver. In this paper, we will focus on the transmitter. The design specifications refer to the SSB radio transceiver with a wavelength of 40 meters which has the following specifications:

- Working frequency: 7MHz
- Output BFO (Beat Frequency Oscillator: 10 MHz
- Output VCO (Variable Controlled Oscillator): 3 MHz
- The mixer, filter and oscillator circuit is bidirectional
- The 1st, 2nd and RF amplifier circuits have a Vbe voltage of about 0.7 volts and a Vce voltage of about 6 volts
- The resulting modulation is SSB (Single Side Band) modulation

Table 1. Radio frequency bands allocation.

| No | Frequency range | Band | Utilization |
|---|---|---|---|
| 1 | ( 3 – 30 ) KHz | VLF (Very Low Frequency) | Navigation, submarine communication |
| 2 | ( 30 – 300 ) KHz | LF (Low Frequency) | Navigation, submarine communication |
| 3 | ( 300 – 3000 ) KHz | MF (Medium Frequency ) | Maritime radio, direction radio, emergency frequency, commercial AM broadcast |
| 4 | ( 3 – 30 ) MHz | HF (High Frequency) | Amateur radio, international broadcasts, long-distance ship and air craft communications, telephone, telegram and faximile |
| 5 | (30 – 300) MHz | VHF (Very High Frequency) | TV broadcast, commercial FM broadcast, AM air craft communication |
| 6 | ( 0.3 – 3 ) GHz | UHF (Ultra High Frequency) | TV broadcasting, navigation, radar, microwave trajectory |
| 7 | ( 3 – 30 ) GHz | SHF (Super High Frequency) | Satellite communications, radar, microwave trajectory |
| 8 | ( 30 – 300 ) GHz | EHF ( Extremely High Frequency) | Satellite radar, 14xperiment and research |

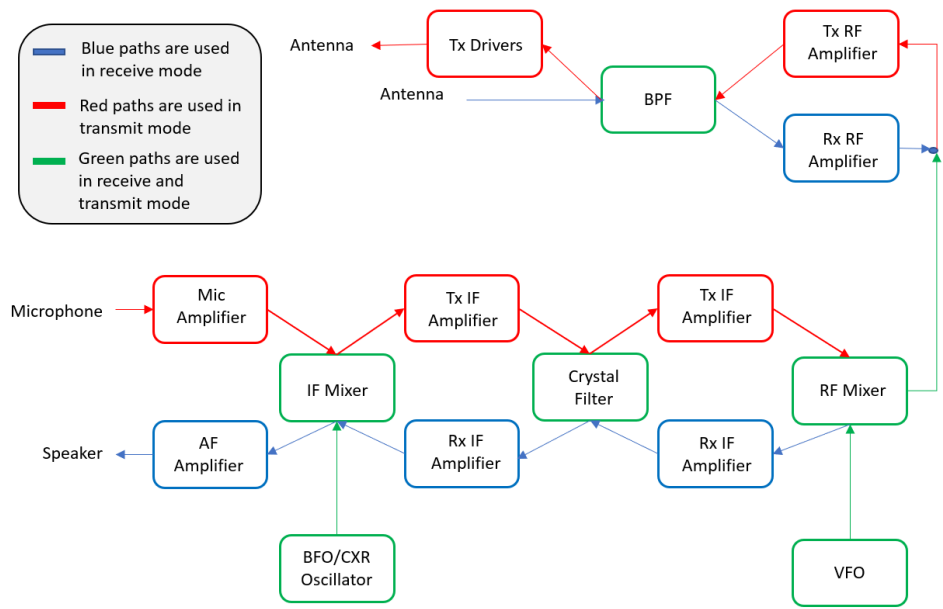

Fig. 1. The diagram block of Bitx radio.

In the first transmitter, the voice signal is amplified by a mic-amplifier, then this signal is modulated with a SBM (Single Balanced Modulator) mixer circuit with the BFO oscillator to the IF frequency. Before entering the SSB (Single Side Band) Filter, the results from the mixer will be amplified by the IF amplifier circuit. After being amplified, the information signal enters the SSB filter to obtain SSB modulation. Then the signal is amplified again by the IF amplifier before entering the mixer circuit. The VFO (Variable Frequency Oscillator) oscillator together with the DBM (Double Balanced Modulator) mixer will produce RF frequencies. Then the signal enters the RF amplifier and is amplified again by the driver amplifier before finally entering the BPF and antenna to be sent. The Bitx circuit block diagram can be seen in Figure 1.

The Figure 1 is a block diagram of the entire Bitx radio transceiver system. The blue color is the part that is used for the receiver mode, the red color is the part that is used for the transmitter mode (transmit) and the green color is used for the second mode, namely the transmitter and the receiver (bidirectional). As explained in the design specifications, we work on the transmitter section which will be discussed in the next sub-chapter

*A. The design of Bitx radio*

In designing the Bitx radio, each circuit block will be described, equipped with a schematic of the circuit, the components used and the design process for each circuit block. The schematic circuit can be found in Figure 2

- **Microphone pre-amplifier**

This microphone amplifier consists of 1 transistor without gain controlled and tone controlled. In designing the mic-amp the author uses a condenser mic with consideration of a low price. This mic amplifier circuit will affect the quality of the resulting modulation. Therefore, make sure that the components used are of good quality.

- **BFO (Beat Frequency Oscillator) and Single Balanced Modulator**

The main function of the BFO together with the SBM is to mix the AF and BFO signals to get the IF signal, on the other hand, to mix the IF and BFO signals to get the AF signal. This BFO is a crystal oscillator equipped with a frequency adjuster, so it can work around its fundamental frequency, this configuration is called VXO (Variable X'tal Oscillator).

The resonant frequency of this BFO must be similar to the SSB filter, in this case the author uses an SSB filter with a center frequency of 10MHz. Thus the BFO should oscillate at 10MHz-1.5KHz = 9.998500Hz and 10MHz+1.5KHz = 10,001,500Hz for LSB and USB reception.

Single Balanced Modulator or SBM functions as a mixer of two signals with different frequencies and produces an output signal whose frequency is the difference between the two frequencies. This modulated signal is formed by two inputs with the condition that the carrier signal must be greater than the voice signal, then mixing occurs and this modulation only passes the information signal. This SBM is made of a triplier coil, two diodes, a variable capacitor and a trimpot as a counterweight. This microphone amplifier consists of 1 transistor without gain controlled and tone controlled. In designing the mic-amp. Trifiller coil is made of ring ferrite or toroid ring. Then it was wound crosswise with 3 strands of 0.3 mm enamel wire twisted 18 cm long by 8 turns.

- **2nd IF Amplifier**

The 2nd IF circuit is a class A amplifier circuit. The configuration of this circuit is very identical to the 1st and RF amplifiers, there are only differences in the value of the components used.

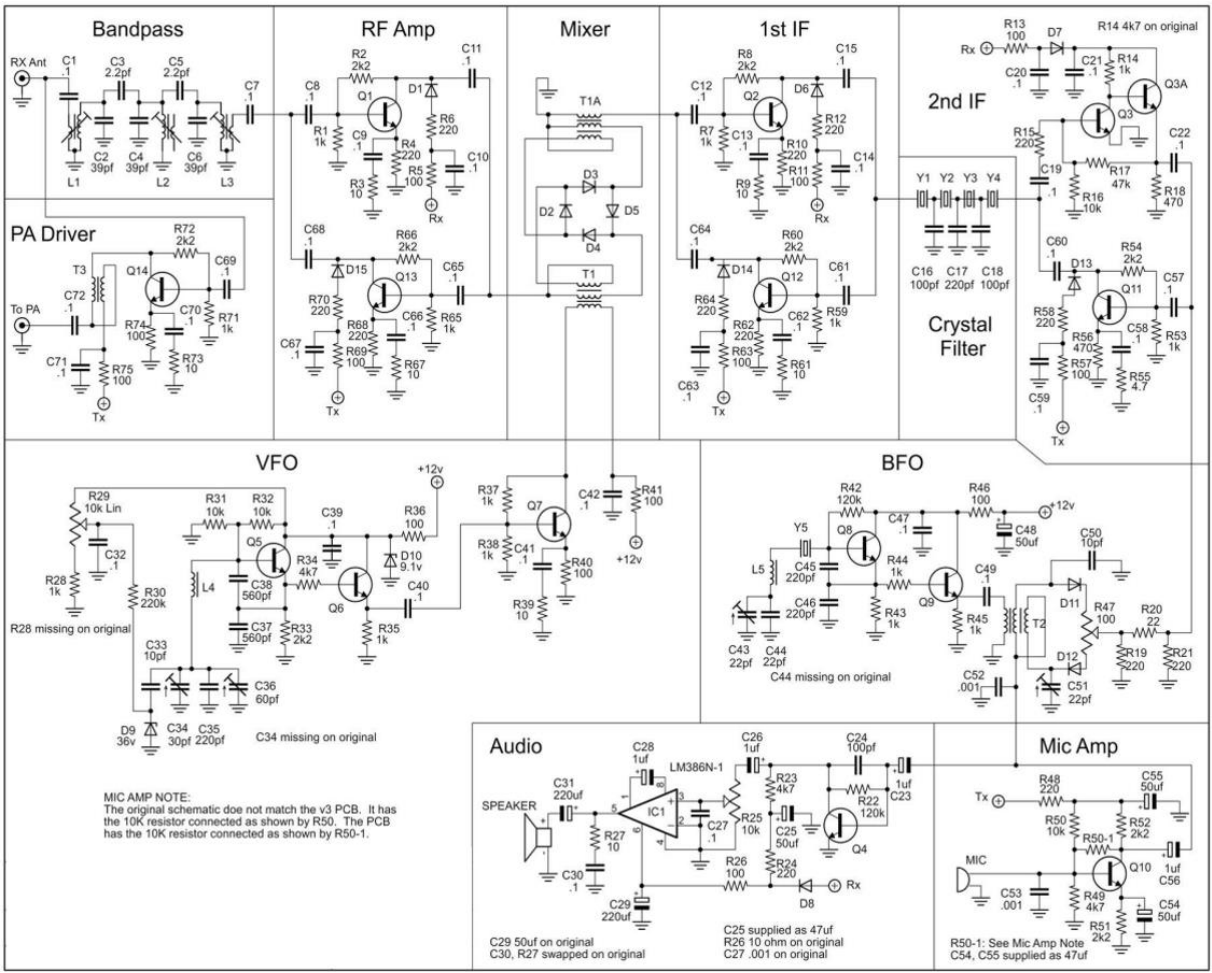

Fig. 2. The diagram block of Bitx radio [1].

Since there is no tuning circuit in this amplifier, the design is quite easy. Make sure that all components are properly installed, especially active components such as transistors.

- **SSB (Single Side Band) Filter**

This SSB filter consists of four crystals with the same frequency arranged in series with some additional components in the form of capacitors. The expected bandwidth is not more than 3 KHz, in accordance with the requirements of radio amateurs for SSB communication (LSB or USB).

- **1st IF Amplifier**

The 1st IF circuit is a class A amplifier circuit. The configuration of this circuit is identical to the 2nd and RF amplifiers, only the difference is in the value of the components used. Since there is no tuning circuit in this amplifier, the design is quite easy. Make sure that all components are properly installed, especially active components such as transistors.

- **VFO (Variable Frequency Oscillator)**

VFO (Variable Frequency Oscillator) is an oscillator in a radio communication system that produces a sine wave as an information signal carrier which is superimposed with a modulation process with adjustable frequency values. VFO as an oscillator with DBM (Double Balanced Modulator) will shift the radio input signal into an IF signal with a frequency around BFO. Thus, the VFO frequency can be searched by the formula:

$$\begin{aligned}
\text{VFO} &= \text{BFO} - \text{Working frequency}, \\
&\text{Since the working frequency} = 7 \text{ MHz, then} \\
\text{VFO} &= \text{BFO} - \text{Working frequency} \\
&= 10 \text{ MHz} - 7 \text{ MHz} \\
&= 3 \text{ MHz}
\end{aligned}$$

The main circuit of the VFO is a transistor whose oscillation frequency is determined by the tank circuit, which is a combination of an inductor and a capacitor. Therefore, the stability of the VFO will be directly proportional to the stability of the value of the capacitor and inductor. So it is attempted to choose a good component and is not susceptible to the influence of temperature, one way is to choose a polystyrene type capacitor, because the polystyrene type capacitor is a type of capacitor that is resistant to the influence of temperature. While the inductor used is made of 0.25 mm diameter enamel wire coils wound on a ferrite core of 18 turns.

- **DBM (Double Balanced Modulator)**

DBM (Double Balanced Modulator) is the same as SBM (Single Balanced Modulator) which functions as a mixer of two signals with different frequencies and produces an output signal whose frequency is the difference between the two frequencies. The difference is with SBM, in DBM the mixed signal is between RF and VFO to produce IF and vice versa mixes IF signal with VFO to produce RF. This DBM circuit consists of two identical trifilar coils and four 1N4148 diodes. Trifiller coil is made of ring ferrite or toroid ring. Then it was wound crosswise with 3 strands of 0.3 mm enamel wire twisted 18 cm long by 8 turns.

- **RF Amplifier**

The RF amplifier circuit is a class A amplifier circuit. The configuration of this circuit is identical to the 2nd and 1st IF amplifiers, only the difference is in the value of the components used. Since there is no tuning circuit in this amplifier, the design is quite easy. Make sure that all components are properly installed, especially active components such as transistors.

- **BPF (Band Pass Filter)**

BPF (Band Pass Filter) consists of three inductors with the same value and several capacitors. The inductor itself is made with a ferrite core coil of 8 mm in diameter and wrapped by enamel wire with a diameter of 0.25 mm. In this BPF, only frequencies that are within the working frequency will be passed. Because the designed Bitx radio works at a wavelength of 40 meters, the frequency that is passed is only about 7 MHz.

*B. Layout PCB and Bitx realization*

The layout PCB of Bitx radio can be seen in Figure 3, while the Bitx realization is shown in Figure 4.

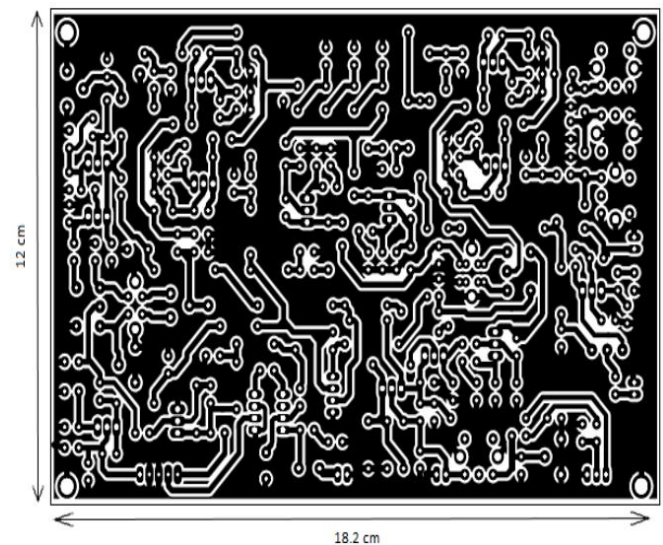

Fig. 1. Design architecture of secure federated learning framework.

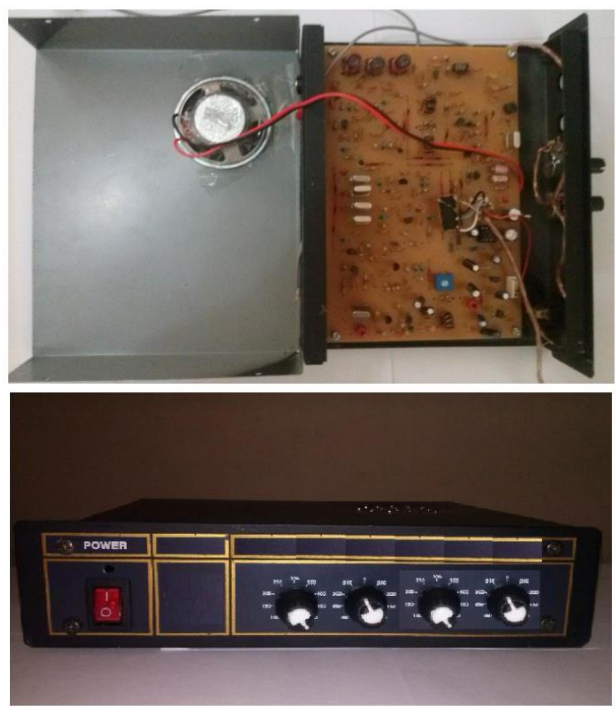

Fig. 3. Realization of radio bidirectional transceiver (Bitx).

IV. PERFORMANCE MEASUREMENT AND NUMERICAL RESULTS

The big concept in the realization of the Bitx radio in this transmitter section is that first the voice signal is amplified by a mic-amplifier, then this signal is modulated with a SBM (Single Balanced Modulator) circuit to an IF frequency of 10 MHz. The resulting signal is still complete both on the USB and LSB sides. However, because the SSB filter has a center frequency of 10 MHz, it must sample the USB signal to obtain the LSB beam. Furthermore, the output signal from the SBM is amplified by an IF amplifier transistor BC 547, before entering the 10 MHz SSB filter. The output of the SSB filter is a signal in the form of SSB modulation. Then the signal is amplified again by the IF amplifier transistor BC 547 before entering the DBM (Double Balanced Modulator) circuit. To get the desired working frequency, the IF signal is then modulated by the DBM with a carrier frequency signal of 3.2 MHz generated from the VFO. The DBM output is then amplified by the RF amplifier and driver amplifier before entering the band pass filter and antenna.

After going through the process of designing and realizing a bidirectional transceiver radio, the next step is to measure the parameters that affect this Bitx radio. Measurements are carried out on each circuit block and will be accompanied by a set-up of measuring tools, equipment used and analysis of tool measurement results. The measurements to be carried out are:

- Measurement of the oscillator circuit with a frequency counter and an oscilloscope
- Measurement of working frequency and output power with spectrum analyzer

- Measurement of the output of each circuit block with a spectrum analyzer.

The equipment used to support the measurement is as follows:
- Oscilloscope GW INSTEK GDS-2104
- Frequency Counter Dynascan Corporation 1805
- Signal Generator HP 8656B 0.9 – 990 MHz
- Spectrum Alanyzer GW INSTEK
- WAVETEK 5070 A Attenuator
- Power Supply GPD 3030
- SANWA YX360TRF Multimeter
- 50 ohm . BNC to BNC connector cable
- Banana plug connector cable 2 pieces

*A. Measurement of the oscillator circuit with a frequency counter and an oscilloscope*

Measurement of the oscillator circuit either BFO or VFO will use two tools, namely frequency counter and oscilloscope. This is to make the measurements more accurate. Figure 4 and Figure 5 shows the result of BFO and VFO measurement using oscilloscope and frequency counter, respectively.

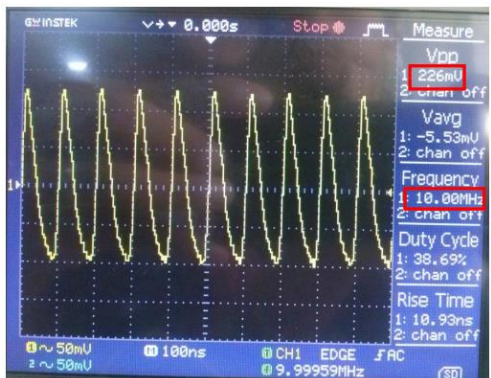

a. BFO (Beat Frequency Oscillator)

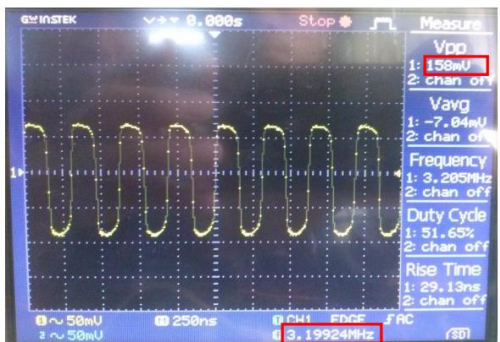

b. VFO (Variable Frequency Oscillator)

Fig. 6. BFO and VFO measurement using oscillator.

From the measurement results based on Figure 4.a it can be seen that the frequency and voltage read on the BFO circuit are 10.00 MHz and 226 mVpp. On the other hand, Based on Figure 4.b, it can be seen that the frequency and voltage read on the VFO circuit are 3.19924 MHz and 158 mVpp.

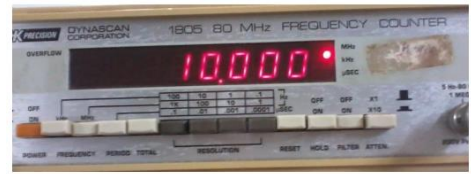

a. BFO (Beat Frequency Oscillator)

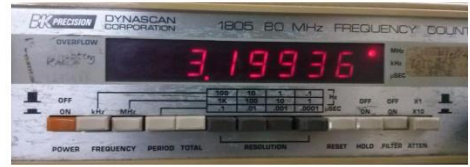

b. BFO (Beat Frequency Oscillator)

Fig. 4. BFO and VFO measurement using frequency counter.

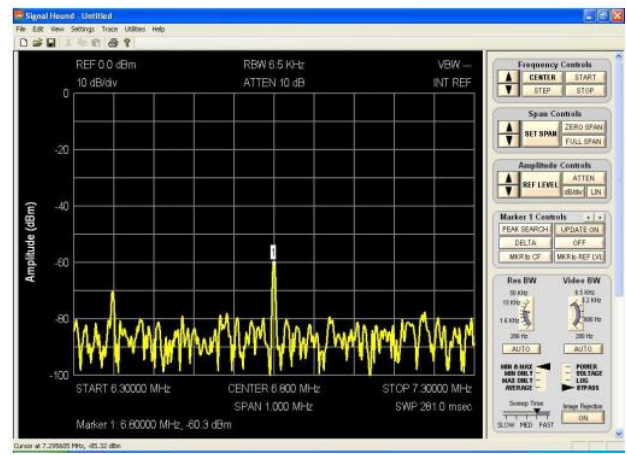

Fig. 5. BFO and VFO measurement using frequency counter.

From the measurement results based on Figure 5.a it can be seen that the frequency read on the BFO circuit is 10 MHz, while based on Figure 42, it can be seen that the frequency read on the VFO circuit is 3.19936 MHz.

*B. Measurement of working frequency and output power with spectrum analyzer*

Based on the measurement results shown in Figure 4 and 5, it can be seen that the read frequencies are 6.79869 MHz and 6.815 MHz. The measurement is almost close to 6.8 MHz according to the working frequency. This working frequency is obtained from the difference between the BFO and VFO frequencies, can be calculated by the formula:

Working frequency = BFO – VFO

= 10 MHz – 3.2 Mhz

= 6.8 MHz

In Figure 6, the measurement of the working frequency using a spectrum analyzer, the results obtained are included with the information signal. In addition, it can also be seen that the figure shows SSB modulation using LSB mode. Meanwhile, the output power of the Bitx radio is very small, only around -50.3 dBm. Therefore we need a power amplifier before entering the antenna.

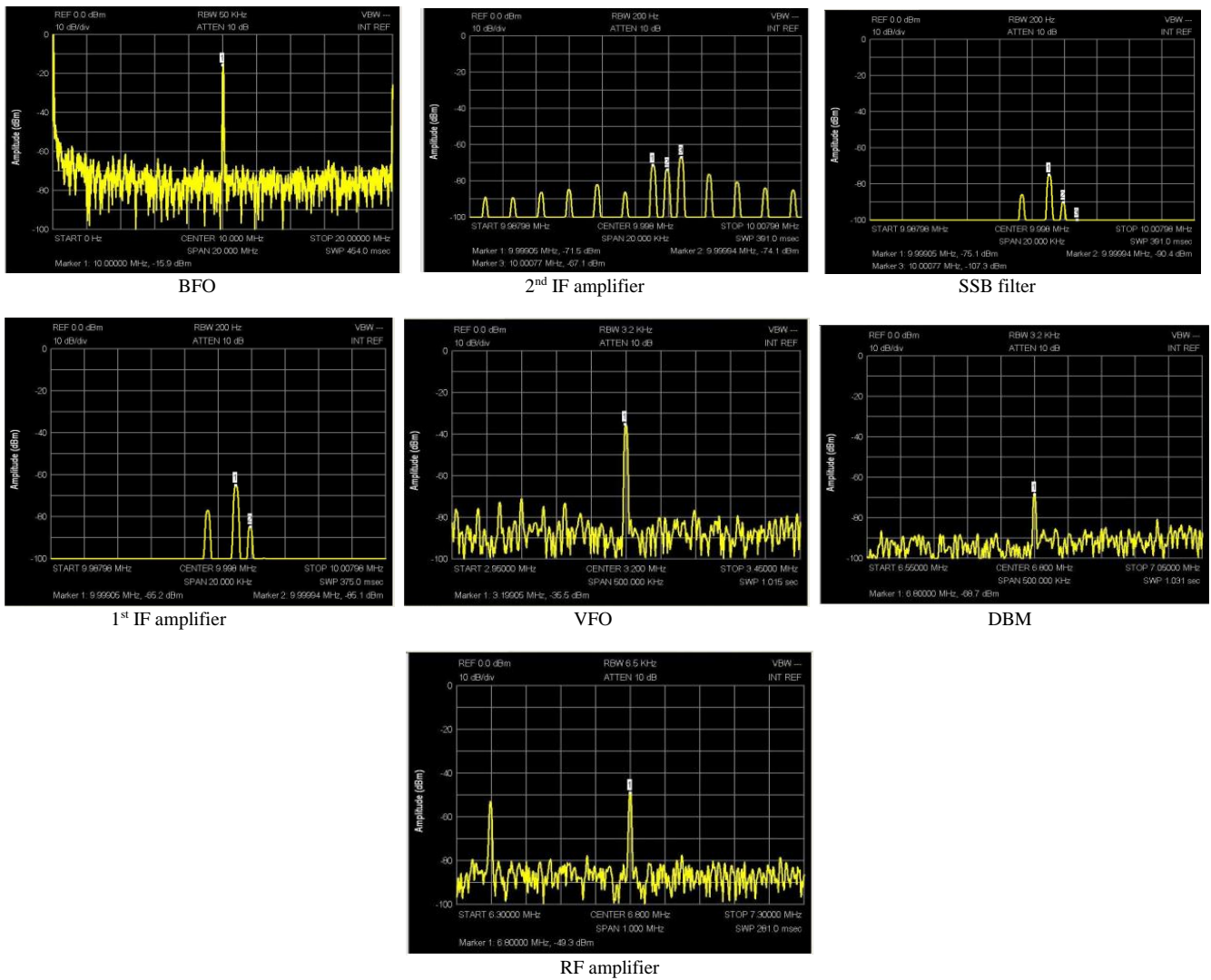

Fig. 7. Block circuits measurement using spectrum analyzer.

*C. Measurement of the output of each circuit block with a spectrum analyzer*

From the measurement results shown in Figure 7, it can be seen that the BFO oscillator is in accordance with the expected results, which is around 10 MHz. The output signal from the 2nd IF amplifier still has a complete LSB and USB signal, while when it enters the SSB filter, the output signal is only an LSB signal. The SSB filter output signal is amplified by the 1st IF amplifier before being mixed by the DBM and VFO. The resulting VFO frequency is almost close to 3.2 MHz, so that the output generated from the DBM is the difference between the IF and the oscillator and the resulting working frequency. Before transmitting the RF signal, it is amplified by the RF amplifier and driver amplifier, until finally the signal is emitted from the Bitx radio.

V. CONCLUSION

Based on the results of the design and realization of a radio bidirectional transceiver (Bitx) in the HF (High Frequency) band, it can be concluded that:

- The resulting working frequency is 6.8 MHz, so it does not match the expected working frequency of 7 MHz. This happens because the VFO oscillator does not produce a frequency of 3 MHz, but 3.2 MHz which is generated from the crystal oscillator. So that the resulting working frequency is 6.8 MHz, which is the difference between the 10 MHz BFO oscillator and the 3.2 MHz VFO oscillator.
- VFO stability is the main key, because it greatly affects the working frequency produced. However, because in this circuit there is often a shift in the working frequency, so in this final project the author replaces the VFO circuit inductor with a 3.2 MHz crystal oscillator to reduce the frequency shift.
- In the resulting output signal, there are still unwanted signals. This is because there is no BPF (Band Pass Filter) which has a function to pass the desired signal and filter out unwanted signals. However, because the power from the Bitx radio is very small, in practice BPF will lower the signal level, so the author eliminates the function of BPF. However, the impact that occurs is that many unwanted signals enter such as harmonic signals and other signals outside the working frequency.